\documentclass[12pt]{JHEP3}
\usepackage{psfrag,amsmath,epsfig,feynarts,cite,latexsym}
\def\dg#1{\frac{\delta\Gamma}{\delta#1}}

\def\pslash#1{{\setbox0=\hbox{$#1$}
  \rlap{\ifdim\wd0>.7em\kern.22\wd0\else\kern.1\wd0\fi /}#1}}
\def\psl{\pslash p}
\def\ksl{\pslash k}

\def\Dsl{\pslash D}

\def\glui{{\tilde{g}}}
\def\epsbar{{\bar{\epsilon}}}

\def\ghat{{\hat{g}}}
\def\gtilde{{\tilde{g}}}
\def\gbar{{\bar{g}}}
\def\gtiltil{{\tilde{\tilde{g}}}}

\allowdisplaybreaks[1]
\preprint{IPPP/05/06\\
DCPT/05/12\\
hep-ph/0503129}
\title{Regularization by Dimensional Reduction: Consistency,
  Quantum Action Principle, and Supersymmetry}
\author{Dominik St\"ockinger
\\  Institute for Particle Physics Phenomenology, Physics Department, 
University of Durham,
Durham DH1~3LE, UK
\\ E-mail: \email{Dominik.Stockinger@durham.ac.uk}%
}
\abstract{
It is proven by explicit construction that regularization by
dimensional reduction can be formulated in a mathematically consistent
way. In this formulation the quantum action principle is shown to
hold. This provides an intuitive and elegant relation between the
$D$-dimensional Lagrangian and Ward or Slavnov-Taylor identities, and
it can be used in particular to study to what extent dimensional
reduction preserves supersymmetry. We give several examples of
previously unchecked cases. 
}

\begin{document}

\section{Introduction}

Regularization of loop contributions has been a notorious
and long standing problem for supersymmetric gauge theories. 
Dimensional regularization (DREG) \cite{HV} and its variant
dimensional reduction (DRED) \cite{Siegel79} are by far the most
common regularization schemes. However, both schemes have well-known
serious shortcomings.

DREG explicitly breaks supersymmetry because the number
of degrees of freedom of gauge bosons and gauginos does not match for
$D\ne4$. In practical calculations this breaking would have to be
corrected by adding suitable supersymmetry-restoring counterterms
\cite{BHZ96,STIChecks}, whose existence is always guaranteed by the
renormalizability of supersymmetric gauge theories
\cite{SSTI1,SSTI2,SSTIus}. But since these counterterms do not
originate from multiplicative renormalization their evaluation and
implementation poses a significant practical complication.

In DRED only momenta are treated in $D$ dimensions whereas
$\gamma$-matrices and gauge fields remain ordinary 4-vectors, and thus
the direct supersymmetry breaking of DREG is avoided. Hence DRED is
used in most 
practical cases where the supersymmetry breaking of DREG would
matter. However, DRED is mathematically inconsistent
\cite{Siegel80}. In \cite{Siegel80} it has been shown that the
combination of 4-dimensional $\gamma$-matrices and $\epsilon$-tensors
with a $D$-dimensional $g^{\mu\nu}$ leads to the relation
$0=D(D-1)(D-2)(D-3)(D-4)$, which is obviously inconsistent with
non-integer values of $D$.

This problem of regularization is not only a curiosity of
supersymmetric field theories. It is of increasing practical
importance as more and more predictions have to be evaluated at the
loop-level in order to match the current or expected future
experimental accuracy. For example, for observables such as the
lightest Higgs boson mass, $(g-2)$ of the muon, or electroweak
precision observables even two-loop corrections have to be known
\cite{HHW}. The need for a better understanding of DRED and the
related $\overline{DR}$ scheme has also been emphasized by the SPA
``Supersymmetry Parameter Analysis'' Project \cite{SPA}.

Recently, interest in properties of DRED has increased also because of
the mass factorization problem observed in Ref.\ \cite{BKNS}
and emphasized again in Ref.\ \cite{NS}. However, in the context of
supersymmetry DRED is most useful for loop diagrams, while a
factorization problem has only been found for real gluon radiation
diagrams. Moreover, many important calculations such as the ones  
reviewed in Ref.\ \cite{HHW} or the evaluation of $\beta$-functions
\cite{JJK} do not rely on factorization at all. In the following we
will focus on the first problem of mathematical inconsistency and
related questions.

Of course, the mathematical inconsistency should
not matter in many practical cases \cite{JJReview}. 
However, this inconsistency (which implies that one initial expression
can lead to different results depending on the calculational steps)
makes it very hard to prove generally valid statements on DRED.

Most importantly, even in DRED it is unclear whether or to what extent
supersymmetry is actually preserved. 
It is only known that some supersymmetry relations are satisfied at
the one- or two-loop level, but even at the one-loop level the checks
do not exhaust all Green functions that could be affected by a
supersymmetry breaking.

In this paper we show the validity of two statements:
\begin{enumerate}
\item DRED can be formulated in a mathematically consistent way
  (Section \ref{sec:DREDFormulation}). The consistent formulation is 
  essentially identical to the way DRED is usually applied, with the
  exception that no Fierz identities can be used. We give an explicit
  representation of all formally $D$-dimensional quantities in terms
  of mathematically well-defined objects. This ensures consistency of
  the calculational prescriptions, and uniqueness and existence of the
  results.
\item In this formulation of DRED, the quantum action principle is
  valid (Section \ref{sec:QAP}). This theorem provides a direct
  relation between 
  the regularized Lagrangian ${\cal L}$ and Ward or
  Slavnov-Taylor identities for Green functions at all orders:
\begin{equation}
i\,\delta\langle T\phi_1\ldots\phi_n\rangle
= \langle T\phi_1\ldots\phi_n\Delta\rangle\ ,
\label{QAP}
\end{equation}
where $\delta=\delta\phi_i\frac{\delta}{\delta\phi_i}$ denotes some
  symmetry transformation and $\Delta=\int d^Dx \,\delta{\cal L}$ (see
  Sec.\ \ref{sec:QAP} for further variants of the theorem). 
  Thus this relation constitutes a simple and
  general tool to study the symmetry-properties of DRED.
\end{enumerate}

In a third step the quantum action principle is applied to
supersymmetry transformations (Section \ref{sec:Application}).
Supersymmetry Ward or Slavnov-Taylor identities 
can be brought into the form
\begin{equation}
\delta_{\rm SUSY}\langle T\phi_1\ldots\phi_n\rangle = 0.
\label{SymId}
\end{equation}
Hence if $\delta_{\rm SUSY}{\cal L}$ were zero, the quantum action
principle would immediately show the validity of supersymmetry Ward
and Slavnov-Taylor identities to all orders. 

However, it turns out that $\delta_{\rm SUSY}{\cal L}\ne0$ in
consistent DRED because in order to avoid the mathematical
inconsistency a ``quasi-4-dimensional'' space (Q4S) \cite{ACV} is
introduced, where Fierz identities do not hold. 
Correspondingly, the supersymmetry identities (\ref{SymId}) 
are not generally valid but only to the extent that the insertion of
$\delta_{\rm SUSY}{\cal L}$ does not contribute on the right-hand side
of eq.\ (\ref{QAP}). 

It is often dramatically simpler to prove that
the right-hand side of (\ref{QAP}) vanishes than to evaluate the
left-hand side of (\ref{SymId}). 
Hence by using the quantum action principle many more Ward or
Slavnov-Taylor identities are amenable to checks, and the checks
themselves are drastically simplified. In Sec.\
\ref{sec:Application} we will apply this strategy and check several
identities of practical interest at the one- and two-loop level.

\section{Consistent dimensional reduction}
\label{sec:DREDFormulation}

In DRED, momenta are continued from 4 to $D$ dimensions, while gauge
fields and $\gamma$-matrices remain 4-dimensional objects. The
$D$-dimensional momenta live in a $D$-dimensional space with
metric tensor $\ghat^{\mu\nu}$, and the 4-dimensional quantities live
in a 4-dimensional space with metric tensor $g^{\mu\nu}$. It is
important that the relation 
\begin{equation}
\psl\psl=\frac12 p_\mu p_\nu \{\gamma^\mu,\gamma^\nu\}=p_\mu p_\nu
g^{\mu\nu}=p^2
\end{equation}
holds for $D$-dimensional momenta such that e.g.\ a Dirac propagator
$(\psl+m)/(p^2-m^2)$ is indeed the inverse of $(\psl-m)$. This
requires that the $D$-dimensional space is a subspace of the
4-dimensional one and one has the projector relations
\begin{equation}
g^{\mu\nu}\ghat_\nu{}^\rho=\ghat^{\mu\rho}.
\label{Subspace}
\end{equation}
For ordinary DREG, it is well-known that the $D$-dimensional space can
be realized only {\em formally}. Actually it is an infinite
dimensional vector space on which $\ghat^{\mu\nu}$ is defined as a
certain operator with all desired properties that resemble
$D$-dimensional behaviour \cite{Wilson,Collins}. Therefore it seems
difficult for DRED to realize such a formally $D$-dimensional, but
actually  infinite dimensional
space that is at the same time a subspace of the 4-dimensional space.
Indeed it is eq.\ (\ref{Subspace}), combined with several purely
4-dimensional relations between the $\epsilon$-tensor and $g^{\mu\nu}$
that leads to the inconsistency found in \cite{Siegel80}.

The solution is to realize the 4-dimensional space as a
``quasi-4-dimensional'' space Q4S that retains essential
4-dimensional properties but is in fact also infinite
dimensional. Such a space was already postulated in \cite{ACV}, and in
App.\ \ref{app} it is shown that it is indeed possible to construct a
hierarchy of two infinite dimensional spaces Q$D$S$\subset$Q4S such
that Q$D$S is formally $D$-dimensional, Q4S formally 4-dimensional and
Q$D$S is a subspace of Q4S. The complement
of the $D$-dimensional space is an $\epsilon=4-D$-dimensional space
Q$\epsilon$S, Q4S=Q$D$S$\oplus$Q$\epsilon$S. 

In App.\ \ref{app} we also give an explicit construction of all
objects in these spaces that are needed for field theoretical
calculations. This  
explicit construction guarantees that all calculational rules are
consistent and that all calculations lead to unambiguous and
well-defined results. In the following we list the properties and
calculational rules of these objects, the metric tensors,
$\gamma$-matrices, $\gamma_5$, and of charge conjugation.

On Q4S, Q$D$S and Q$\epsilon$S, metric tensors $g^{\mu\nu}$,
$\ghat^{\mu\nu}$ and $\gtilde^{\mu\nu}$ can be defined that satisfy
the following relations:
\begin{subequations}
\label{gmunu}
\begin{align}
g^{\mu\nu} &= \ghat^{\mu\nu}+\gtilde^{\mu\nu} &
g^{\mu\nu}g_{\mu\nu} & =4\\
g^{\mu\nu}\ghat_\nu{}^\rho &=\ghat^{\mu\rho} &
\ghat^{\mu\nu}\ghat_{\mu\nu} &=D \\
g^{\mu\nu}\gtilde_\nu{}^\rho &=\gtilde^{\mu\rho}&
\gtilde^{\mu\nu}\gtilde_{\mu\nu} &=\epsilon=4-D\\
\ghat^{\mu\nu}\gtilde_\nu{}^\rho &=0
\end{align}
\end{subequations}
In practical calculations it is sufficient to know these
relations. The only practical consequence of the infinite dimensional
nature of the underlying spaces is that index counting is not
possible: E.g.\ out of five Lorentz indices $\mu_1$, $\mu_2$, $\mu_3$,
$\mu_4$, $\mu_5$ two must be equal in truly four dimensions, but in
Q4S all of them can be different.

In Q4S, $\gamma$-matrices can be defined that satisfy the Dirac algebra
\begin{align}
\{\gamma^\mu,\gamma^\nu\} &=2g^{\mu\nu}
\label{GammaAlgebra}
\end{align}
and have the hermiticity property
\begin{align}
(\gamma^\mu)^\dagger &=\gamma^0\gamma^\mu\gamma^0.
\end{align}
Again these relations resemble 4-dimensional behaviour and are
sufficient in practice, but as a consequence of Q4S further relations
like Fierz identities are not valid.

Concerning the definition of $\gamma_5$, there are two options. 
Either we define it as in 4 dimensions and as in the HVBM-scheme
\cite{HV,BM} as
$\bar{\gamma}_5=i\gamma^0\gamma^1\gamma^2\gamma^3$, or we define
$\gamma_5$ as totally anticommuting (the existence of such an object
that anticommutes with all infinitely many $\gamma^\mu$ is shown in
App.\ \ref{app}). To distinguish the
two options we denote the purely 4-dimensional one by
$\bar{\gamma}_5$. If the second option is used, we have
\begin{align}
\{\gamma_5,\gamma^\mu\}&=0,
\label{GAnticommuting}
\end{align}
and this is the way DRED is usually defined. Hence the results we will
obtain for this case will immediately hold for all applications of
DRED that rely on eqs.\
(\ref{gmunu},\ref{GammaAlgebra},\ref{GAnticommuting}). 
The drawback of this option is that traces of $\gamma_5$ and four or
more $\gamma$-matrices vanish in contrast to the 4-dimensional case.%
\footnote{The proof is analogous to the DREG-case if we write
  $\gamma^\mu=\hat\gamma^\mu+\tilde{\gamma}^\mu$ with
  $\hat\gamma^\mu\hat\gamma_\mu=D$,
  $\tilde{\gamma}^\mu\tilde{\gamma}_\mu=4-D$. Starting e.g.\ from
${\rm Tr}(\hat\gamma^\alpha\hat\gamma_\alpha\gamma_5
\hat\gamma^\mu\hat\gamma^\nu\hat\gamma^\rho\hat\gamma^\sigma)$ one obtains
$(4-D){\rm Tr}(\gamma_5
\hat\gamma^\mu\hat\gamma^\nu\hat\gamma^\rho\hat\gamma^\sigma)=0$.
  Similarly one obtains $(3-D){\rm Tr}(\gamma_5
\tilde{\gamma}^\mu\hat\gamma^\nu\hat\gamma^\rho\hat\gamma^\sigma)=0$,
and so on. Finally one can show in this way that all traces of the
  form ${\rm Tr}(\gamma_5(\mbox{arbitrary number of }\gamma^{\mu_i}))$
  vanish if $D$ is non-integer.
}
In this respect DRED is not better than DREG with anticommuting
$\gamma_5$. For the 
option of $\bar\gamma_5$ it is useful to introduce additional
projectors onto the true 4-dimensional space, and its complement,
$\gbar^{\mu\nu}$, $\gtiltil^{\mu\nu}$:
\begin{subequations}
\begin{align}
g^{\mu\nu}&=\gbar^{\mu\nu}+\gtiltil^{\mu\nu}&
\gbar^{\mu\nu}\gtiltil_\nu{}^\rho &=0\\
g^{\mu\nu}\gbar_\nu{}^\rho &=\gbar^{\mu\rho} &
\gbar^{\mu\nu}\gbar_{\mu\nu}&=4\\
g^{\mu\nu}\gtiltil_\nu{}^\rho &=\gtiltil^{\mu\rho} &
\gtiltil^{\mu\nu}\gtiltil_{\mu\nu}&=0
\end{align}
\end{subequations}
Using the notation $\tilde{\tilde{a}}^\mu=\gtiltil^{\mu}{}_\nu
a^\nu$, we then have the anticommutation relation
\begin{align}
\{\bar{\gamma}_5,\gamma^\mu\}&=
2\bar\gamma_5 \tilde{\tilde{\gamma}}^{\mu}.
\end{align}
Both definitions of $\gamma_5$ have the additional properties
\begin{subequations}
\begin{align}
\gamma_5^\dagger &=\gamma_5&
\bar{\gamma}_5^\dagger &=\bar{\gamma}_5\\
\gamma_5^2 & = 1&
\bar{\gamma}_5^2 & = 1
\end{align}
\end{subequations}

Particularly for calculations in supersymmetric models, it is very
important to use charge conjugation in order to simplify products of
spinors and $\gamma$-matrices. In App.\ \ref{app} the charge
conjugation matrix $C$ is constructed in Q4S, showing that it is
possible to use all the usual charge conjugation relations even in
Q4S. Charge conjugated spinors $\psi^C$ can be defined and bilinear
expressions can be rewritten in the following way:
\begin{subequations}
\begin{align}
\psi^C& = -i\gamma^0 C \psi^*&
\overline{\psi_1^C} &=-i\psi^T C\\
\bar\psi_1\Gamma\psi_2 &= \overline{\psi_2^C}\Gamma^C\psi_1^C&
\mbox{ with }
\Gamma^C&=-C\Gamma^T C
\end{align}
\end{subequations}
In the rearrangement formula for $\bar\psi_1\Gamma\psi_2$,
anticommuting spinors have been assumed. If $\psi_{1,2}$ commute,
there is an additional minus sign on the right-hand side. The results
for $\Gamma^C$ are
\begin{equation}
\{1,\gamma_5,\bar{\gamma}_5,\gamma^\mu,\gamma^\mu\gamma_5,
\gamma^\mu\bar{\gamma}_5\}^C =
\{1,\gamma_5,\bar{\gamma}_5,-\gamma^\mu,-\gamma_5\gamma^\mu,
-\bar{\gamma}_5\gamma^\mu\}.
\end{equation}

\section{Quantum action principle}
\label{sec:QAP}

The quantum action principle (\ref{QAP}) is a simple relation between
the symmetry properties of the regularized Lagrangian and the full
Green functions. Heuristically, it can be derived in the following way
using the functional integral
\begin{align}
Z(J) &= \int {\cal D}\phi e^{i\int d^Dx({\cal L}+J\phi)},
\end{align}
where a symbolic notation $\phi$ for all quantum fields and $J$ for
all corresponding sources is used. Performing a variable
transformation
\begin{align}
\phi&\to\phi+\delta\phi
\end{align}
in the functional integral and expanding to first order in $\delta$
yields
\begin{align}
0 &=  \int{\cal D}\phi \left(
 \int d^Dx \, i(\delta{\cal L} + J\delta\phi)\right)
e^{i\int d^Dx({\cal L}+J\phi)}.
\end{align}
This equation corresponds exactly to eq.\ (\ref{QAP}) for Green
functions. In addition to transformations of quantum fields, often
variations of external fields or parameters are considered.  For such
cases similar relations hold. In summary, the quantum action
principle has three variants:
\begin{itemize}
\item{Variation of quantum fields:} $\delta=\int d^Dx
  \delta\phi_i(x)\frac{\delta}{\delta\phi_i(x)}$. 
\begin{equation}
i\,\delta\langle T\phi_1\ldots\phi_n\rangle
= \langle T\phi_1\ldots\phi_n\Delta\rangle\ ,
\end{equation}
where $\Delta=\int d^Dx \,\delta{\cal L}$ and $\delta$ has to be
  pulled into the brackets on the left-hand   side.
\item{Variation of an external (non-propagating) field $Y(x)$:}
\begin{equation}
-i\,\frac{\delta}{\delta Y(x)}\langle T\phi_1\ldots\phi_n\rangle
= \langle T\phi_1\ldots\phi_n\Delta\rangle\ ,
\end{equation}
with $\Delta=\frac{\delta}{\delta Y(x)}\int d^Dx {\cal L}$.
\item{Variation of a parameter $\lambda$:}
\begin{equation}
-i\frac{\partial}{\partial\lambda}
\langle T\phi_1\ldots\phi_n\rangle
= \langle T\phi_1\ldots\phi_n\Delta\rangle\ ,
\end{equation}
with $\Delta=\frac{\partial}{\partial\lambda}\int d^Dx {\cal L}$.
\end{itemize}
The two other variants can be similarly derived from the functional
integral, and the different signs can be easily understood in this
way.

However, the derivation from the functional integral is only heuristic
because the measure ${\cal D}\phi$ has tacitly been assumed to be
invariant under 
the variable transformation. But this is precisely the point where the
regularization enters. Hence the quantum action principle has to be
established separately for each regularization.

For BPHZ-renormalization, it has been proven in \cite{QAPBPHZ}, and
for DREG, it has been proven in \cite{BM}. The proofs for the three
variants are very similar, so we restrict ourselves here to the first
one. We will see that the proof for DREG can be applied with no
essential modification also to DRED. A crucial prerequisite however is
the mathematical consistency 
of DRED as formulated in Sec.\ \ref{sec:DREDFormulation}.

Partly, the proof of eq.\ (\ref{QAP}) is pure combinatorics of Feynman
diagrams, and this part is independent of the regularization and not
discussed here. The more intricate part of the proof is related to the
fact that Feynman diagrams treat the kinetic part of the Lagrangian
different from the interaction part. Decomposing ${\cal L}+J\phi={\cal
  L}_0+{\cal L}_{\rm int}$, where ${\cal L}_0$ contains the kinetic
terms determining the propagators and ${\cal L}_{\rm int}$ contains
the interaction and source terms, the generating functional $Z(J)$ has
the following expression in terms of Feynman diagrams:
\begin{align}
Z(J) &= 
\left\langle T
\exp(i
{\textstyle\int} d^Dx {\cal L}_{\rm int})
\right\rangle .
\end{align}
This expression has to be evaluated with free fields using Wick's
theorem to obtain the contractions giving rise to the propagators.
Similarly, the Feynman diagram identity corresponding to eq.\
(\ref{QAP}) is given by
\begin{align}
0 &= \left\langle T(\textstyle\int d^Dx 
\big(\delta{\cal L}_0+\delta{\cal L}_{\rm int}\big)\big)
\exp(i\textstyle\int d^Dx {\cal L}_{\rm int})\right\rangle.
\label{QAPDiagrams}
\end{align}
Hence in order to prove this identity one has to show that the terms
involving $\delta {\cal L}_0$ cancel the ones involving $\delta
{\cal L}_{\rm int}$. Writing ${\cal L}_0=\frac12 \phi_i D_{ij}\phi_j$
with some differential operator $D_{ij}$, the terms involving
$\delta{\cal L}_0=\delta\phi_i D_{ij}\phi_j$ result in two different
kinds of Wick contractions involving $\phi_j$:
\begin{align}
 &\left\langle \big(\textstyle\int d^Dx
 \begin{array}[t]{c}
\delta\phi_i\\\uparrow
\end{array} D_{ij}
\begin{array}[t]{c}\phi_j\\\uparrow\end{array}\big)
 \big(i\textstyle\int d^Dx{\cal L}_{\rm int}\big)\ldots
 \big(i\textstyle\int d^Dx{\cal L}_{\rm int}\big)\right\rangle
\nonumber\\
+&
\left\langle \big(\textstyle\int d^Dx\,\delta\phi_i\, D_{ij}
\begin{array}[t]{c}\phi_j\\\uparrow\end{array}\big)
\big(i\textstyle\int d^Dx\begin{array}[t]{c}{\cal L}_{\rm int}\\
  \uparrow\end{array}\big)
\big(i\textstyle\int d^Dx{\cal L}_{\rm int}\big)\ldots
\big(i\textstyle\int d^Dx{\cal L}_{\rm int}\big)\right\rangle,
\label{Wickcontractions}
\end{align}
where the Wick contractions involving $\phi_j$ are denoted by
arrows (Wick contractions of all other fields are not made explicit
here). In momentum space 
Feynman diagrams, the first line always produces a loop integral over
$D_{ij}$ times the propagator $P_{jk}$ from $\phi_j$ to some field
$\phi_k$ within the composite operator $\delta\phi_i$. Now one crucial
property of DRED and DREG is relevant, namely
\begin{align}
D_{ij}P_{jk}&=i\delta_{ik}
\label{Propagators}
\end{align}
in momentum space even on the regularized level. Hence the loop
integral corresponding to the first contraction in
(\ref{Wickcontractions}) is scaleless and therefore zero.

Contracting $\phi_j$ in the second line of (\ref{Wickcontractions})
with some field $\phi_k$ within ${\cal L}_{\rm int}$ results in the
product $P_{jk} \frac{\delta{\cal L}_{\rm int}}{\delta \phi_k}$. Using
eq.\ (\ref{Propagators}) again, we find that eq.\
(\ref{Wickcontractions}) becomes
\begin{align}
i^2\left\langle \big(\textstyle\int d^Dx\, \delta\phi_i \frac{\delta{\cal
    L}_{\rm int}}{\delta\phi_i}\big)
\big(i\textstyle\int d^Dx{\cal L}_{\rm int}\big)\ldots
\big(i\textstyle\int d^Dx{\cal L}_{\rm int}\big)\right\rangle.
\end{align}
Using this result in eq.\ (\ref{QAPDiagrams})  together with
combinatorial arguments shows that the terms involving $\delta {\cal
  L}_0$ indeed cancel the ones involving $\delta {\cal L}_{\rm int}$,
proving the quantum action principle.

To summarize, the two main points where the properties of DRED (shared
by DREG) enter are 
eq.\ (\ref{Propagators}) and the fact that scaleless integrals vanish
in dimensional schemes. In \cite{BM}, eq.\ (\ref{Propagators}) has
been proven not only in momentum space but also using the
$\alpha$-representation for the propagators, where the relation
is less obvious. The same proof is also valid in DRED
without changes. However it relies on the fact that the expression
$(\psl+m)/(p^2-m^2)$ for a Dirac propagator is really the inverse of
the kinetic operator $(\psl-m)$, even if $p$ is $D$-dimensional and
the $\gamma$-matrices are defined in Q4S. The importance of this fact
and its relation to the mathematical inconsistency of DRED with truly
4-dimensional $\gamma$-matrices has already been stressed in Sec.\
\ref{sec:DREDFormulation}.

\section{Supersymmetry of DRED}
\label{sec:Application}

In this section we apply the quantum action principle to obtain more
information on the supersymmetry properties of DRED. What is known so
far is limited to many one-loop cases and few cases beyond one-loop
order, and the methods used up to now are not very easy to apply to
complicated situations since they involve the evaluation of full Green
functions including the finite parts. The application of the quantum
action principle should lead to a significant simplification, and we
will illustrate this with several non-trivial examples. We will start
however with a brief overview of the current status (see also the
review \cite{JJReview}).

\subsection{Previous checks}

Immediately after the invention of DRED \cite{Siegel79}, first checks
of Ward identities were performed in \cite{CJN80}. The checks
comprised several gauge Ward identities up to the two-loop level and
supersymmetry Ward identities at the one-loop level. However, only
supersymmetry Ward identities for propagators but not for 3- or
4-point functions were considered. In \cite{ACV} the quantum action
principle was assumed and it was argued that the propagator Ward
identity in a supersymmetric gauge theory without matter fields should
hold up to the 
three-loop level. In \cite{BHZ96} an S-matrix identity connecting the
gluino-squark-quark 3-point function and gauge interactions was
checked at the one-loop level. 

Ref.\ \cite{BM85} demonstrated that supersymmetry Ward identities are
not sufficient to describe the symmetry content of supersymmetric
gauge theories if (as done in virtually all practical cases) the
Wess-Zumino gauge is used. Instead, supersymmetry Slavnov-Taylor
identities \cite{SSTI1,SSTI2,SSTIus} constitute an exact
representation of the symmetry content 
of these theories. They are the center of proofs of the absence of
anomalies and the multiplicative renormalizability of supersymmetric
gauge theories. In Refs.\ \cite{STIChecks} several Slavnov-Taylor
identities relating propagators, 3-point functions, and Green
functions corresponding to loop-corrected supersymmetry
transformations have been checked.

In all checked cases, the respective identities turned out to be valid
in DRED on the regularized level. However, it is noteworthy that even
at the one-loop level there are Green functions such as four-scalar
interactions that do not appear in any of the checked identities and
hence could in principle still violate supersymmetry in DRED. In
addition to the direct checks, in several cases DRED-results for
$\beta$-functions of supersymmetric parameters could be checked either
for internal consistency or against non-perturbative results
\cite{betachecks}. 

\subsection{General breaking of the Slavnov-Taylor identity in DRED}

The present situation of DRED and supersymmetry is not completely
satisfactory: So far DRED has passed all checks and there are many
reasons to believe that DRED has an 
even wider range of validity, but there is no general proof or
analysis whether or to what extent DRED really preserves
supersymmetry. The quantum action principle is a suitable tool to
perform such a general analysis. In the following we show how it can
be applied to study supersymmetry Slavnov-Taylor identities in DRED.

The Slavnov-Taylor identities can be combined into a single equation
for the generating functional $\Gamma$ for the one-particle
irreducible (1PI) vertex functions:
\begin{align}
S(\Gamma)&=0
\label{STI}
\end{align}
with
\begin{align}
S(\Gamma)&=\int d^Dx\bigg(
\sum_i\dg{Y_i(x)}\dg{\varphi_i(x)}+\sum_j s\varphi'_j\dg{\varphi'_j}
\bigg).
\end{align}
Here $\varphi_i$ denote the quantum fields with non-linear BRS
transformations, $Y_i$ the corresponding external fields that couple
to these BRS transformations, and $\varphi'_j$ denote the fields with
linear BRS transformations $s\varphi'_j$ (see Refs.\
\cite{SSTI1,SSTI2,SSTIus} for 
further details). Eq.\ (\ref{STI}) describes the full symmetry
content, i.e.\ gauge invariance, supersymmetry and translational
invariance, and it should hold for the {\em renormalized} vertex
functional $\Gamma$. The interesting question is whether it already
holds for $\Gamma^{\rm DRED}$, the {\em regularized} vertex functional
in DRED where no counterterms have been added.

By applying the usual Legendre transformations, we can apply the
quantum action principle to the one-particle irreducible vertex
functions and obtain
\begin{align}
S(\Gamma^{\rm DRED}) &=
i[S(\Gamma_{\rm cl})]\cdot \Gamma^{\rm DRED},
\label{STIQAP}
\end{align}
where $[X]\cdot \Gamma^{\rm DRED}$ denotes the insertion of an
operator $X$ into the 1PI vertex functions analogous to the insertion
on the right-hand side of eq.\ (\ref{QAP}). $\Gamma_{\rm cl}$ denotes
the tree-level contribution which is nothing but the classical action
$\int d^Dx {\cal L}$. 

Eq.\ (\ref{STIQAP}) can be used as the master equation for the study
of symmetry properties of DRED. In order to make use of it, we will
evaluate $S(\Gamma_{\rm cl})$ for a general supersymmetric gauge
theory. For simplicity we assume a simple gauge group with gauge field
$A^\mu_a$, gaugino $\tilde{g}_a$, and chiral matter multiplets
$(\phi_i,P_L\psi_i)$ with the projector $P_L=\frac12(1-\gamma_5)$ and
the gauge group generators $T^a_{ij}$. The superpotential and its
derivatives are denoted as $W(\phi)$, $W_i=\partial W/\partial\phi_i$,
etc. We
use the general conventions of Ref.\ \cite{SSTIus}, rewritten in terms
of 4-spinors in order to make use of the calculational rules 
introduced in Sec.\ \ref{sec:DREDFormulation}. In particular, the
supersymmetry ghost is a commuting Majorana spinor
$\epsilon=\epsilon^C$, the Faddeev-Popov ghost is denoted as $c_a$ and
the sources for the BRS transformations $s\varphi_i$ generically  as
$Y_{\varphi_i}$. In the following formulas we write
$\tilde{g}_{ij}=\tilde{g}_a T^a_{ij}$ and suppress indices wherever
possible. Using the anticommuting version of $\gamma_5$, we obtain
\begin{subequations}
\label{DeltaRes}
\begin{align}
S(\Gamma_{\rm cl})&=
\int d^Dx\left[\Delta_{\rm gauge}+\Delta_{\rm fix}+\Delta_{\rm matter}+
\Delta_{\rm pot}+\Delta_{{Y_\glui}} + \Delta_{Y_c}
+\Delta_{{Y_\psi}}\right],
\\
\Delta_{\rm gauge} &=
\frac{i}{2}gf_{abc}(\epsbar\gamma^\mu\glui_a)
(\overline{\glui}_b\gamma_\mu\glui_c),
\\
\Delta_{\rm fix} &=
0,
\\
\Delta_{\rm matter} &=
-g\Big[
2(\overline{\psi} P_R\epsilon)(\overline{\glui}P_L\psi)
+2(\epsbar P_L\psi_j)(\overline{\psi}_i P_R\glui_{ij})
+(\overline{\psi}_i\gamma^\mu P_L\psi_j)(\epsbar\gamma_\mu \glui_{ij})
\Big],
\\
\Delta_{\rm pot} &=
-\frac{1}{\sqrt2}(\overline\psi^C_i P_L \psi_j)
(\epsbar P_L \psi_k) W_{ijk} +h.c.,
\\
\Delta_{{Y_\glui}} &=
-(\overline{Y_{\glui_a}}\sigma^{\mu\nu}\epsilon)
(D_\mu \epsbar\gamma_\nu\glui_a)
-i(\epsbar\gamma^\mu\epsilon)(\overline{Y_{\glui_a}}D_\mu\glui_a)
-i(\epsbar\gamma_5\gamma^\mu D_\mu\glui_a)
(\overline{Y_{\glui_a}}\gamma_5\epsilon),
\\
\Delta_{Y_c} &=
iY_{c_a} (\epsbar\gamma^\mu\epsilon)(\epsbar\gamma_\mu\glui_a),
\\
\Delta_{{Y_\psi}} &=
-\sqrt2 g\Big[ 
2(\overline{Y_{\psi}}P_L\epsilon)(\epsbar P_R\glui)\phi
+(\epsbar\gamma^\mu\glui_{ij})
 (\overline{Y_{\psi_i}}P_L\gamma_\mu\epsilon)\phi_j\Big]
\nonumber\\
&\phantom{=}{}+2i\Big[
(\overline{Y_{\psi}}P_L\epsilon)\epsbar\gamma^\mu 
+(\overline{Y_{\psi}}P_L \gamma^\mu\epsilon)\epsbar
-\frac12(\epsbar\gamma^\mu\epsilon)\overline{Y_{\psi}} \Big]
D_\mu P_L\psi
+h.c.
\end{align}
\end{subequations}
All these terms vanish identically in four dimensions where Fierz
rearrangements are possible. The term $\Delta_{\rm gauge}$ was already
obtained in \cite{ACV} in the discussion of supersymmetry Ward
identities. As will be shown in the next subsection, the insertion of
these terms into diagrams as in (\ref{STIQAP}) still vanishes in many
cases, which explains the fact that no supersymmetry-violation of DRED
has been found so far.

If the strictly 4-dimensional option $\bar\gamma_5$ is used in the
definition of the regularized Lagrangian, the regularized theory and
the breaking $S(\Gamma_{\rm cl})$ is modified. For example,
$\Delta_{\rm fix}$ and $\Delta_{\rm matter}$ contain the additional
terms
\begin{align}
&\bar{c}_a\epsbar\tilde{\tilde{\gamma}}_\mu\bar\gamma_5\epsilon\partial^\mu
  D_a
-\sqrt2\Big[
\phi^\dagger\epsbar\Dsl\tilde{\tilde{\Dsl}}P_L\psi
+\overline{\psi}P_R\tilde{\tilde{\Dsl}}\Dsl\epsilon\phi
-i
F^\dagger\epsbar \tilde{\tilde{\Dsl}}P_L\psi
-i
\overline{\psi}P_R\tilde{\tilde{\Dsl}}\epsilon F\Big],
\end{align}
where $D_a=-g\phi^\dagger T^a\phi$ and $F^\dagger_i=-\partial
W/\partial\phi_i$ with the superpotential $W$. It can be easily seen
that inserting these terms into the simplest Green functions such as
$\langle T\phi^\dagger\psi \Delta_{\rm matter}\rangle$ leads to
non-vanishing contributions already at the one-loop level. Hence if
$\bar\gamma_5$ is used instead of $\gamma_5$, even the simplest
Slavnov-Taylor identities relating one-loop propagators are violated,
just like in DREG. Therefore and since in all practical applications
of DRED the fully anticommuting $\gamma_5$ is used we will restrict
ourselves to this case in the following.

\subsection{Examples}
 
Now we use the master equation (\ref{STIQAP}) and the explicit result
(\ref{DeltaRes}) for the breaking to study the supersymmetry
properties of DRED. We focus on three examples of practical interest
but will exhibit also several general features. The examples are%
\footnote{In order not to overload the following formulas, we suppress
  all indices that are not summed over. Furthermore, we always write
  just $\psi$ instead of $P_L\psi$ for the fermionic components of
  chiral multiplets.}
\begin{itemize}
\item The Slavnov-Taylor identity relating the $\phi$ and $\psi$
  propagators $\Gamma_{\phi^\dagger\phi}$ and
  $\Gamma_{\psi\overline{\psi}}$. This identity is derived from the
  derivative  
$\frac{\delta^3 S(\Gamma)}{\delta\phi^\dagger\delta\psi\delta\epsbar}
=0$
of (\ref{STI}) and reads
\begin{align}
\label{STI1}
0 &=
\Gamma_{\psi \epsbar {Y_{\phi}}_i}\Gamma_{\phi^\dagger\phi_i}-
\Gamma_{\phi^\dagger {Y_{\psi}}_i\epsbar}\Gamma_{\psi\overline{\psi}_i}.
\end{align}
This identity expresses in particular the mass
degeneracy of the bosonic and fermionic components of a chiral
multiplet and is obviously one of the most
fundamental supersymmetry relations.
\item The Slavnov-Taylor identity relating the loop-corrected
  supersymmetry transformations of $\phi$ and $\psi$,  
$\Gamma_{Y_\phi \psi\epsbar}$ and
$\Gamma_{\phi\epsilon \overline{Y}_\psi}$. It is derived from
  $\frac{\delta^4 S(\Gamma)}{\delta
  \phi\delta\epsilon\delta\epsbar\delta Y_\phi}=0$
and reads
\begin{align}
\label{STI2}
0&=
\Gamma_{Y_\phi \psi_i\epsbar}
\Gamma_{\phi\epsilon {\overline{Y}_\psi}_i}
-\Gamma_{\phi  {Y_\psi^C}_i\epsbar}
\Gamma_{Y_{\phi}\epsilon\,\overline{\psi}^C_i}
+2\psl
+\ldots
\end{align}
where $p$ is the momentum flowing into $\phi$ and the dots denote terms
involving power-counting finite Green functions that vanish at
tree-level and do not receive counterterm contributions.
The appearance of loop corrections to the supersymmetry
transformations is discussed in \cite{STIChecks,SiboldSusyTrans}. The
above identity expresses the fact that the loop-corrected
supersymmetry transformations still have to be in agreement with the
supersymmetry algebra $\{Q,\bar{Q}\}=2\psl$.
\item
The Slavnov-Taylor identity determining the $\phi^4$ interaction. In
supersymmetric models the coupling constant of a $\phi^4$ interaction
is never a free parameter but it can be related to gauge and
superpotential couplings. For example, in the Minimal Supersymmetric
Standard Model, the quartic Higgs boson self coupling can be entirely
expressed in terms of the gauge coupling. This is of paramount
importance for phenomenology since it leads to the tree-level
prediction $M_h<M_Z$ and to strong constraints on the Higgs boson mass
$M_h$ even at the loop level (see \cite{HHW} and references therein).
The corresponding Slavnov-Taylor identity can be derived from
$\frac{\delta^5 S(\Gamma)}
{\delta\phi^\dagger\delta\phi\delta\phi^\dagger
\delta\psi\delta\epsbar}=0$ and reads
\begin{align}
\label{STI3}
0&=
\Gamma_{\psi\epsbar {Y_{\phi}}_i}
\Gamma_{\phi^\dagger\phi\phi^\dagger\phi_i}
+2\Gamma_{\phi\phi^\dagger {Y_{\glui}}_a\epsbar}
\Gamma_{\phi^\dagger\psi\overline{\glui}_a}
-\Gamma_{\phi^\dagger\phi^\dagger {Y_{\psi_i}^C}\epsbar}
\Gamma_{\phi\psi\overline{\psi}^C_i}
+\ldots
\end{align}
It relates $\Gamma_{\phi^\dagger\phi\phi^\dagger\phi_i}$ to a gauge
interaction (second term) and a superpotential interaction (third
term). Such an identity has not been considered in the literature so
far.
\end{itemize}
On the regularized level, the identities (\ref{STI1}), (\ref{STI2}),
(\ref{STI3}) are not necessarily satisfied. The right-hand side
adds up to an expression that can be computed using the quantum action
principle in the form (\ref{STIQAP}). For example, for the first
identity we obtain
\begin{align}
\Gamma^{\rm DRED}_{\psi \epsbar {Y_{\phi}}_i}
\Gamma^{\rm DRED}_{\phi^\dagger\phi_i}-
\Gamma^{\rm DRED}_{\phi^\dagger {Y_{\psi}}_i\epsbar}
\Gamma^{\rm DRED}_{\psi\overline{\psi}_i}
&=i
([S(\Gamma_{\rm cl})]\cdot\Gamma^{\rm DRED})_{\phi^\dagger\psi\epsbar}
,
\end{align}
where the right-hand side denotes the 1PI Green function with external
$\phi^\dagger$, $\psi$, $\epsbar$ and an insertion of $S(\Gamma_{\rm
  cl})$, eq.\ (\ref{DeltaRes}). Whether such Green functions vanish or
constitute a supersymmetry breaking of DRED can often be determined
quite easily by inspection of the corresponding Feynman diagrams.

All insertions in (\ref{DeltaRes}) are four-fermion
operators, and the Green functions in our examples always have one
external $\epsbar$, a second external fermion line and one internal
fermion loop. So all Feynman diagrams we have to consider involve one
of the basic fermion topologies in Fig.\ \ref{fig:basic} to which
additional boson lines are attached. Topology (a) corresponds to an
insertion of $\Delta_{\rm gauge}$; Topologies (b) and
(c$_{\glui}$) correspond to insertions 
of $\Delta_{\rm matter}$ where either the gaugino or $\psi$ acts as
the external fermion. Topology (c$_\psi$) originates from an
insertion of $\Delta_{\rm pot}$. Denoting the string 
of $\gamma$-matrices 
attached to the second external fermion line as $A$ and the
$\gamma$-string associated to the closed fermion loop as $B$, the
Feynman rules lead to the following general results:
\begin{subequations}
\label{ResFormulas}
\begin{align}
\label{ResA}
\mbox{Topology \parbox{3em}{(a):} }\quad&
\gamma^\mu (B-B^C)\gamma_\mu A
+\gamma^\mu A {\rm Tr}(\gamma_\mu B)
,\\
\label{ResB}
\mbox{Topology \parbox{3em}{(b):} }\quad&
2(P_L B P_R - P_R B^C P_L)A
 - \gamma^\mu A {\rm Tr}(\gamma_\mu P_L B),
\\
\label{ResC}
\mbox{Topology \parbox{3em}{(c$_\glui$):} }\quad&
(\gamma^\mu B \gamma_\mu -2 P_R B^C )P_L A
- 2 P_L A {\rm Tr}(P_R B),
\\
\label{ResCpot}
\mbox{Topology \parbox{3em}{(c$_\psi$):} }\quad&
P_L (B+B^C) P_L A - P_L A {\rm Tr}(P_L B).
\end{align}
\end{subequations}
All these expressions vanish identically in strictly four dimensions.
In DRED and Q4S, however, it can be explicitly checked that the
expressions only vanish if $B$ does not contain more than 4, 2, or 3
$\gamma$-matrices in case (a), (b), or (c), respectively. 
E.g.\ if we set $B=\gamma^{\mu_1}\gamma^{\mu_2}\gamma^{\mu_3}$  in
case (b) we obtain
\begin{align}
2P_L\left(
\gamma^{\mu_1}\gamma^{\mu_2}\gamma^{\mu_3}
-\gamma^{\mu_1}g^{\mu_2\mu_3}+\gamma^{\mu_2}g^{\mu_1\mu_3}
-\gamma^{\mu_3}g^{\mu_1\mu_2}
\right)
\end{align}
instead of zero, so this expression could lead to a violation of a
Slavnov-Taylor identity if it appears in an actual diagram. 

\FIGURE[ht]{
\unitlength=1.cm%
\begin{feynartspicture}(14,4)(3,1)
\FADiagram{Topology (a)}
\FAVert(10,16){1}
\FALabel(3,20)[r]{$\epsbar$\ }
\FAProp(17,20)(10,16)(0.,){/Straight}{0}
\FALabel(17,20)[l]{\ ${\glui}$}
\FAProp(3,20)(10,16)(0.,){/Straight}{0}
\FALabel(4,14)[r]{$\glui$}
\FALabel(16,14)[l]{$\glui$}
\FAProp(10,16)(10,4)(1.,){/Straight}{0}
\FAProp(10,16)(10,4)(-1.,){/Straight}{0}
\FADiagram{Topology (b)}
\FAVert(10,16){1}
\FALabel(3,20)[r]{$\epsbar$\ }
\FAProp(17,20)(10,16)(0.,){/Straight}{0}
\FALabel(17,20)[l]{\ ${\glui}$}
\FAProp(3,20)(10,16)(0.,){/Straight}{0}
\FALabel(4,14)[r]{$\psi$}
\FALabel(16,14)[l]{$\overline{\psi}$}
\FAProp(10,16)(10,4)(1.,){/Straight}{0}
\FAProp(10,16)(10,4)(-1.,){/Straight}{0}
\FADiagram{Topology (c$_{\glui,\psi}$)}
\FAVert(10,16){1}
\FALabel(3,20)[r]{$\epsbar$\ }
\FAProp(17,20)(10,16)(0.,){/Straight}{0}
\FALabel(17,20)[l]{\ ${\psi}$}
\FAProp(3,20)(10,16)(0.,){/Straight}{0}
\FALabel(4,14)[r]{$\glui$,$\psi$}
\FALabel(16,14)[l]{$\overline{\psi}$}
\FAProp(10,16)(10,4)(1.,){/Straight}{0}
\FAProp(10,16)(10,4)(-1.,){/Straight}{0}
\FAVert(5.75,5.75){0}
\FALabel(2,3)[r]{${\phi^\dagger}$\ }
\FAProp(5.75,5.75)(2,3)(0.,){/ScalarDash}{1}
\end{feynartspicture}
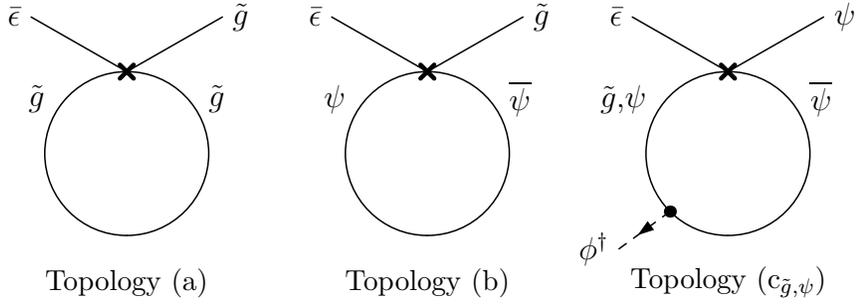
\caption{Basic topologies of diagrams involving an insertion of
  $S(\Gamma_{\rm cl})$, eq.\ (\ref{DeltaRes}). Only fermion lines are
  drawn; additional boson lines have to be attached in the actual
  diagrams. The index $f$ of Topology (c$_f$) corresponds to the
  fermion $f=\glui,\psi$ appearing in the loop.
  In the text, the $\gamma$-string attached to the second
  external fermion line is denoted as $A$, the $\gamma$-string
  attached to the closed fermion loop as $B$.}
\label{fig:basic}
}

\FIGURE[ht]{
\parbox{3cm}{\ }
\unitlength=1.cm%
\begin{feynartspicture}(5,4)(1,1)
\FADiagram{}
\FAVert(10,16){1}
\FALabel(3,20)[r]{$\epsbar$\ }
\FAProp(17,20)(10,16)(0.,){/Straight}{0}
\FALabel(17,20)[l]{\ ${\psi}$}
\FAProp(3,20)(10,16)(0.,){/Straight}{0}
\FALabel(4,14)[r]{$\glui$,$\psi$}
\FALabel(16,14)[l]{$\overline{\psi}$}
\FAProp(10,16)(10,6)(1.,){/Straight}{0}
\FAProp(10,16)(10,6)(-1.,){/Straight}{0}
\FAVert(6.46,7.46){0}
\FALabel(1,4)[r]{${\phi^\dagger}$\ }
\FAProp(6.46,7.46)(1,4)(0.,){/ScalarDash}{1}
\FAVert(13.54,7.46){0}
\FALabel(19,4)[l]{\ ${\phi^\dagger}$}
\FAProp(13.54,7.46)(19,4)(0.,){/ScalarDash}{1}
\FAVert(10,6){0}
\FALabel(10,1)[t]{${\phi}$}
\FAProp(10,6)(10,1)(0.,){/ScalarDash}{-1}
\end{feynartspicture}
\parbox{3cm}{\ }
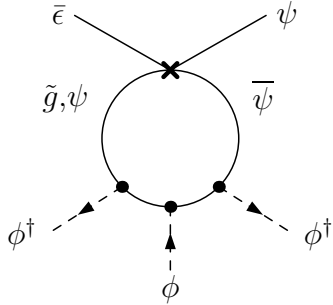
\caption{One-loop diagram corresponding to
$([S(\Gamma_{\rm cl})]\cdot\Gamma^{\rm DRED})_{
\phi^\dagger\phi^\dagger
\phi\psi\epsbar}$.}
\label{fig:phi4}
}

Now let us discuss the $\phi^4$-identity (\ref{STI3}) at the one-loop
level. Its possible violation is given by the diagram in Fig.\
\ref{fig:phi4} (+permutations of the boson lines). After integrating
over the loop momentum, the $\gamma$-string $B$ can only contain the
covariants $\psl_1$, $\psl_2$, $\psl_3$ with the three independent
external momenta of the diagram. Hence after suitable
simplifications $B$ cannot contain more than
three $\gamma$-matrices, and this is not enough to lead to a
non-vanishing contribution. This shows that identity (\ref{STI3}) is
valid in DRED at the one-loop level. By the same token, however, it is
possible that (\ref{STI3}) is violated at the two-loop level where
e.g.\ the exchange of an additional vector boson can lead to more
covariants within $B$.

Secondly we turn to identity (\ref{STI1}) relating the $\phi$ and
$\psi$ propagators. At the one-loop level, it has been verified
already in \cite{STIChecks}   in various  
supersymmetric models by explicit evaluation of all Feynman diagrams
corresponding to the Green functions in (\ref{STI1}). The same
verification is almost trivial if the quantum 
action principle is used: The only one-loop diagram
contributing to the violation $([S(\Gamma_{\rm cl})]\cdot
\Gamma^{\rm DRED})_{\phi^\dagger\psi\epsbar}$ is the ``Topology (c)''
diagram in Fig.\ \ref{fig:basic}. In this diagram, the $\gamma$-string
$B$ contains at most two $\gamma$-matrices, and so the expression
(\ref{ResC}) or (\ref{ResCpot}) and thus the whole diagram vanishes.

At the two-loop level there are several diagrams; Fig.\ \ref{fig:prop}
shows one of them. After integrating over the fermion loop momentum
and contracting all indices within the $\gamma$-string $B$, $B$ can
only contain the covariants $\psl$, $\ksl$, where $k$ is the second
loop momentum, and $\gamma^\mu$ if $A$ and $B$ are connected by a
vector boson as in Fig.\ \ref{fig:prop}. Hence $B$ can contain at most
three $\gamma$-matrices if there is a virtual vector boson (which
implies that the diagram is based on Topology (c)) and at most
two $\gamma$-matrices in all other cases. Thus in all cases the
respective expressions (\ref{ResFormulas})
vanish. This shows that the propagator identity (\ref{STI1}) is valid
in DRED even at the two-loop level.

\FIGURE[ht]{
\parbox{3cm}{\ }
\unitlength=1.cm%
\begin{feynartspicture}(3.5,3.5)(1,1)
\FADiagram{}
\FAVert(10,16){1}
\FALabel(3,20)[r]{$\epsbar$\ }
\FAProp(19,20)(10,16)(0.,){/Straight}{0}
\FALabel(19,20)[l]{\ ${\psi}$}
\FAProp(3,20)(10,16)(0.,){/Straight}{0}
\FALabel(4,14)[r]{$\glui$,$\psi$}
\FALabel(13.5,12.5)[r]{$\overline{\psi}$}
\FAProp(10,16)(10,4)(1.,){/Straight}{0}
\FAProp(10,16)(10,4)(-1.,){/Straight}{0}
\FAVert(5.75,5.75){0}
\FALabel(2,3)[r]{${\phi^\dagger}$\ }
\FAProp(5.75,5.75)(2,3)(0.,){/ScalarDash}{1}
\FAVert(16,18.66){0}
\FAVert(16,10){0}
\FAProp(16,10)(16,18.66)(.7,){/Sine}{0}
\end{feynartspicture}
\parbox{3cm}{\ }
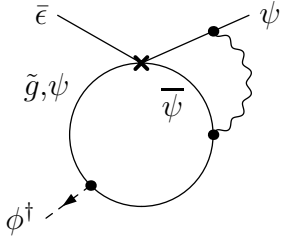
\caption{A two-loop diagram corresponding to 
$([S(\Gamma_{\rm cl})]\cdot\Gamma^{\rm
    DRED})_{\phi^\dagger\psi\epsbar}$. 
Such two-loop
  diagrams involving a virtual vector boson are the only ones where
the $\gamma$-string $B$ can contain three $\gamma$-matrices after
  integration over the fermion loop momentum.}
\label{fig:prop}
}

Similar arguments can be applied to identity (\ref{STI2}) for the
supersymmetry transformations. There is no one-loop diagram
contributing to $([S(\Gamma_{\rm cl})]\cdot
\Gamma^{\rm DRED})_{\phi\epsilon\epsbar Y_\phi}$, which immediately
proves that (\ref{STI2}) is valid in DRED at the one-loop level.
The two-loop diagrams are shown in Fig.\ \ref{fig:susy}. After
integrating over the fermion loop momentum, the $\gamma$-string $B$
can only contain the covariants $\psl$ and $\ksl$, and thus $B$ cannot
contain more than two $\gamma$-matrices. Thus again in all cases the
respective expressions (\ref{ResFormulas})
vanish, and (\ref{STI2}) holds in DRED at the two-loop level. 

\FIGURE[ht]{
\unitlength=1.cm%
\begin{feynartspicture}(3.5,3.5)(1,1)
\FADiagram{}
\FAVert(7,16){1}
\FALabel(1,20)[r]{$\epsbar$\ }
\FAProp(17,16)(7,16)(0.2,){/Straight}{0}
\FAProp(17,16)(20,20)(0.,){/Straight}{0}
\FAVert(17,16){0}
\FALabel(20,20)[l]{\ ${\epsilon}$}
\FAProp(1,20)(7,16)(0.,){/Straight}{0}
\FALabel(13,18)[b]{$\glui$}
\FALabel(1,11)[r]{$\glui$}
\FAProp(7,16)(7,6)(1.,){/Straight}{0}
\FAProp(7,16)(7,6)(-1.,){/Straight}{0}
\FAVert(3.46,7.46){0}
\FALabel(1,3)[r]{${\phi^\dagger}$\ }
\FAProp(3.46,7.46)(1,3)(0.,){/ScalarDash}{1}
\FAVert(7,6){0}
\FALabel(13,4)[t]{$\phi$}
\FAProp(7,6)(17,6)(0.2,){/ScalarDash}{-1}
\FAVert(17,6){0}
\FALabel(20,3)[l]{\ $Y_\phi$}
\FAProp(17,6)(20,3)(0.,){/ScalarDash}{-1}
\FALabel(20,11)[l]{$c$}
\FAProp(17,16)(17,6)(-0.4,){/GhostDash}{1}
\end{feynartspicture}
\qquad
\begin{feynartspicture}(3.5,3.5)(1,1)
\FADiagram{}
\FAVert(7,16){1}
\FALabel(1,20)[r]{$\epsbar$\ }
\FAProp(17,16)(7,16)(0.2,){/Straight}{0}
\FAProp(17,16)(20,20)(0.,){/Straight}{0}
\FAVert(17,16){0}
\FALabel(20,20)[l]{\ ${\epsilon}$}
\FAProp(1,20)(7,16)(0.,){/Straight}{0}
\FALabel(13,18)[b]{$\glui$}
\FALabel(1,11)[r]{$\psi$}
\FAProp(7,16)(7,6)(1.,){/Straight}{0}
\FAProp(7,16)(7,6)(-1.,){/Straight}{0}
\FAVert(3.46,7.46){0}
\FALabel(1,3)[r]{${\phi^\dagger}$\ }
\FAProp(3.46,7.46)(1,3)(0.,){/ScalarDash}{1}
\FAVert(7,6){0}
\FALabel(13,4)[t]{$\phi$}
\FAProp(7,6)(17,6)(0.2,){/ScalarDash}{-1}
\FAVert(17,6){0}
\FALabel(20,3)[l]{\ $Y_\phi$}
\FAProp(17,6)(20,3)(0.,){/ScalarDash}{-1}
\FALabel(20,11)[l]{$c$}
\FAProp(17,16)(17,6)(-0.4,){/GhostDash}{1}
\end{feynartspicture}
\qquad
\begin{feynartspicture}(3.5,3.5)(1,1)
\FADiagram{}
\FAVert(7,16){1}
\FALabel(1,20)[r]{$\epsbar$\ }
\FAProp(17,16)(7,16)(0.2,){/Straight}{0}
\FAProp(17,16)(20,20)(0.,){/Straight}{0}
\FAVert(17,16){0}
\FALabel(20,20)[l]{\ ${\epsilon}$}
\FAProp(1,20)(7,16)(0.,){/Straight}{0}
\FALabel(15,17.5)[b]{$\glui$}
\FALabel(9,17.5)[b]{$\psi$}
\FALabel(1,11)[r]{$\glui$,$\psi$}
\FAProp(7,16)(7,6)(1.,){/Straight}{0}
\FAProp(7,16)(7,6)(-1.,){/Straight}{0}
\FAVert(12.,17){0}
\FALabel(12.,21)[b]{${\phi^\dagger}$}
\FAProp(12.,17)(12.,21)(0.,){/ScalarDash}{1}
\FAVert(7,6){0}
\FALabel(13,4)[t]{$\phi$}
\FAProp(7,6)(17,6)(0.2,){/ScalarDash}{-1}
\FAVert(17,6){0}
\FALabel(20,3)[l]{\ $Y_\phi$}
\FAProp(17,6)(20,3)(0.,){/ScalarDash}{-1}
\FALabel(20,11)[l]{$c$}
\FAProp(17,16)(17,6)(-0.4,){/GhostDash}{1}
\end{feynartspicture}
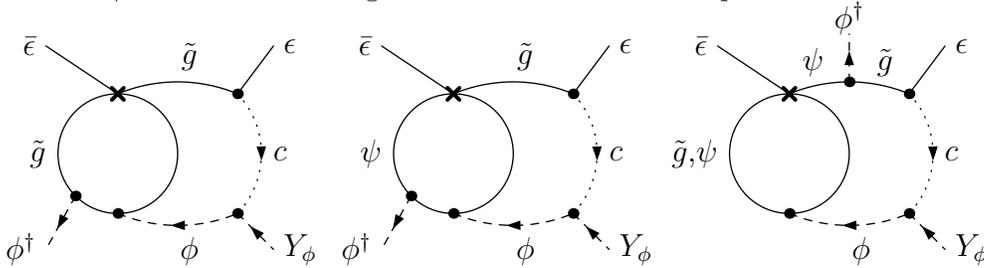
\caption{Two-loop diagrams corresponding to $([S(\Gamma_{\rm
      cl})]\cdot\Gamma^{\rm DRED})_{\phi\epsilon\epsbar Y_\phi}$.}
\label{fig:susy}
}

\section{Conclusions}

In this paper we have tried to put DRED on a firm mathematical basis
by explicitly constructing the formally $D$- and 4-dimensional objects
and by establishing the quantum action principle. If the calculational
rules of Section \ref{sec:DREDFormulation} are applied, no internal 
contradictions can appear. The quantum action principle provides an
elegant all-order relation between the symmetry-properties of the
Lagrangian $\delta_{\rm SUSY}{\cal L}$ and Ward or
Slavnov-Taylor identities.

Using this relation we have studied the supersymmetry-properties of
DRED.  If a totally anticommuting $\gamma_5$ is used as
traditionally done in DRED, the breaking term (\ref{DeltaRes})
consists only of four-fermion operators that would vanish if Fierz
identities were valid. We have shown that there are simple rules as to
when these expressions can be non-vanishing, depending only on the
number of $\gamma$-matrices in the closed fermion line. These rules
can be easily applied to study the breaking of Slavnov-Taylor
identities; hence by using the quantum action principle many more
identities are amenable to checks, and the checks themselves are
drastically simplified.


Here we considered the identities for propagators and
supersymmetry transformations of matter fields. It was straightforward
to show that these identities are valid up to the two-loop level in
DRED. In addition, we verified the identity for the
$\phi^4$-interaction, which had not been considered in the literature
before, at the one-loop level. The rules for the insertions indicate
that at higher orders supersymmetry relations could be violated. In
that case, supersymmetry-restoring counterterms might be necessary in
DRED similar to DREG, but to a much smaller extent. 

It will be an
important future task to delineate the precise range of validity of
the supersymmetry relations in DRED and to find the necessary
counterterms. Furthermore, it would be desirable to include soft
supersymmetry breaking into the analysis along the lines of Refs.\
\cite{SSTIus}. However, given our examples it seems plausible that for
many foreseeable calculations up to the two-loop level DRED can be used
without having to worry about breaking of supersymmetry.

The treatment of $\gamma_5$ still causes problems. Like naive
DREG, DRED with totally anticommuting $\gamma_5$ necessarily implies
that traces involving $\gamma_5$ and an arbitrary number of
$\gamma^\mu$'s are zero. Hence e.g.\ fermion triangle diagrams with an
axial vector current cannot be correctly computed in this scheme. The
alternative definition of a strictly 4-dimensional $\gamma_5$ avoids
this defect, but like in the HVBM-scheme of DREG, supersymmetry and
gauge invariance are broken already at one-loop order.

However, in any given calculation it is usually easy to see whether
this problem 
is relevant, and very often it is not (see e.g.\ Ref.\ \cite{g-2} for
the discussion of an example at the two-loop level). Moreover, for the
$\gamma_5$-problem in DREG there are elaborate 
practical prescriptions, e.g.\ in Refs.\ \cite{Gamma5}.
Using the present results as a basis one can hope to find similar
prescriptions for supersymmetric theories within DRED that yield
correct results for traces but allow the use of the anticommuting
$\gamma_5$ and avoid breaking of supersymmetry and gauge invariance as
much as possible. 

\subsection*{Acknowledgements}

I thank the participants of the SPA-workshop on DRED at DESY, January
2005, where this work was begun, for stimulating discussions. In
particular I am grateful to W.~Hollik, I.~Jack, P.~Zerwas, and 
M.~Gorbahn for many discussions, questions and comments.

\begin{appendix}
\section{Explicit construction of Q4S}
\label{app}

In this Appendix we give an explicit construction of the objects
introduced in Sec.\ \ref{sec:DREDFormulation} and show that they
satisfy all desired relations. First, the momentum integral is
obviously identical in DRED and DREG. In \cite{Wilson} it has been
shown that a formally $d$-dimensional integral of functions of the form
$f(p^2,pq_1,\ldots pq_n)$ can be defined if the underlying vector
space is infinite dimensional. $d$-dimensionality of the integral is
expressed in terms of the scaling and normalization properties
\begin{align}
\int d^dp f(sp)&=s^{-d}\int d^dp f(p),&
\int d^dp e^{-p^2/2} &= \pi^{d/2}.
\end{align}
While the contravariant metric tensor $g^{\mu\nu}$ on this infinite
dimensional vector space is defined by its components, its covariant
counterpart $g_{\mu\nu}$ is defined as a bilinear form on this space
as
\begin{align}
g_{\mu\nu}a^{\mu\nu} &\equiv
\frac{d\Gamma(d/2)}{\pi^{d/2}}\int d^dp
\, p_\mu p_\nu a^{\mu\nu} \delta(p^2-1).
\end{align}
With this definition one has the formally $d$-dimensional result
$g_{\mu\nu}g^{\mu\nu}=d$. Further properties of the $d$-dimensional
integral and metric tensor are discussed in \cite{Collins}. 

In DRED, we simply perform the construction of the $d$-dimensional
integral and metric tensor twice: once with $d=D$ (and metric
signature $+--\ldots$), and once with
$d=\epsilon=4-D$ (and signature $---\ldots$). This yields two infinite
dimensional vector spaces Q$D$S and Q$\epsilon$S
and two metric tensors $\ghat^{\mu\nu}$ and $\gtilde^{\mu\nu}$ with
$\ghat_{\mu\nu}\ghat^{\mu\nu}=D$,
$\gtilde_{\mu\nu}\gtilde^{\mu\nu}=\epsilon$.  The $D$-dimensional
integral in the first of these spaces constitutes the momentum
integral in DRED, the $\epsilon$-dimensional integral of Q$\epsilon$S
has no other purpose than defining $\gtilde_{\mu\nu}$. We can 
then construct the quasi-4-dimensional space Q4S as the direct sum Q4S=Q$D$S$\oplus$Q$\epsilon$S and define its metric tensor as
$g^{\mu\nu}\equiv \ghat^{\mu\nu}+\gtilde^{\mu\nu}$.%
\footnote{
Any vector $p$ in Q4S has the decomposition
$p={\hat{p}\choose\tilde{p}}$, where $\hat{p}\in$Q$D$S,
$\tilde{p}\in$Q$\epsilon$S. We can naturally identify
$\hat{p}\equiv{\hat{p}\choose0}$ and 
$\tilde{p}\equiv{0\choose\tilde{p}}$ and write
$p^\mu=\hat{p}^\mu+\tilde{p}^\mu$. Similarly, the relation
$g^{\mu\nu}\equiv \ghat^{\mu\nu}+\gtilde^{\mu\nu}$ really means the
decomposition $g\equiv{\ghat\ 0\choose0\ \gtilde}$. 
Strictly speaking, the indices $\mu$ in Q4S can be considered as {\em
  pairs} $\mu=(a;i)$, where $a=1,2$ and $i=0,1,2,3,\ldots$ such 
that $p^{(1;i)}=\hat{p}^i$, $p^{(2;i)}=\tilde{p}^i$.
}
Then all equations (\ref{gmunu}) are satisfied  by construction. 

In order to define $\gamma$-matrices in Q4S, we rely on the existence
of $\gamma$-matrices in DREG, in particular on the the explicit
representation of Ref.\ \cite{Collins}. Denoting these matrices as
$\Gamma^\mu$ ($\mu=0,1,2\ldots\infty$), they satisfy
$\{\Gamma^\mu,\Gamma^\nu\}=2g^{\mu\nu}$,
$\Gamma^\mu{}^\dagger=\Gamma^0\Gamma^\mu\Gamma^0$, and after suitable
reordering $\Gamma^\mu{}^*=(-1)^\mu\Gamma^\mu$. In DRED, according to
the structure of Q4S as the sum of two vector spaces, the vector of
$\gamma$-matrices has the decomposition
$\gamma={\hat\gamma\choose\tilde{\gamma}}$ or 
$\gamma^\mu=\hat\gamma^\mu+\tilde{\gamma}^\mu$.
We define
\begin{align}
\hat\gamma^0
&=
\left(\begin{array}{cc}\Gamma^0&0\\0&\Gamma^0\end{array}\right),&
\hat\gamma^i
&=
\left(\begin{array}{cc}\Gamma^{2i+4}&0\\0&\Gamma^{2i+4}
\end{array}\right)\quad(i\ge1),\\
\tilde{\gamma}^\mu
&=
\left(\begin{array}{cc}0&-i\Gamma^{2\mu+5}\\i\Gamma^{2\mu+5}&0
\end{array}\right)\quad(\mu\ge0),\\
\gamma_5
&=
\left(\begin{array}{cc}0&\Gamma^{4}\\-\Gamma^{4}&0
\end{array}\right),&
C
&=
\left(\begin{array}{cc}0&-\Gamma^{2}\Gamma^0\\\Gamma^{2}\Gamma^0&0
\end{array}\right).
\end{align}
One can immediately verify that this definition satisfies all
properties listed in Sec.\ \ref{sec:DREDFormulation}, and in addition
we can read off the properties
\begin{align}
\gamma^\mu{}^*&=\gamma^\mu, &
\gamma_5^* &= \gamma_5, &
C^* &= C,\\
C^2&=-1 , &
\{C,\gamma^0\}&=0, &
C^\dagger &= -C,
\end{align}
which implies in particular $(\psi^C)^C=\psi$ for any spinor.

Traces of a product $A$ of $\gamma$-matrices can be defined by
\cite{Collins} 
\begin{align}
{\rm Tr}A &\equiv \lim_{N\to\infty}\frac{4}{N}\sum_{j=1}^N A_{jj},
\end{align}
so that in particular ${\rm Tr}1=4$. Since all traces of
$\gamma$-matrices can be reduced to ${\rm Tr}1$ times metric tensors
by applying the anticommutation relations, results of traces are
formally the same in four dimensions and in Q4S as long as no
$\gamma_5$ is present. 

\end{appendix}

\begin{flushleft}

\end{flushleft}


\begin{thebibliography}{99} 


\bibitem{HV} G.~'t Hooft and M.~Veltman,
               {\em Nucl. Phys.} {\bf B 44} (1972) 189.

\bibitem{Siegel79} W.~Siegel,
                       {\em Phys. Lett.} {\bf B 84} (1979) 193.

\bibitem{BHZ96}
W.~Beenakker, R.~H\"opker and P.~M.~Zerwas,
{\em Phys.\ Lett.}{\bf\ B} {\bf 378} (1996) 159
[arXiv:hep-ph/9602378].

\bibitem{STIChecks}
W.~Hollik, E.~Kraus and D.~St\"ockinger,
{\em Eur.\ Phys.\ J.}{\bf\ C} {\bf 11} (1999) 365
[arXiv:hep-ph/9907393];\\
W.~Hollik and D.~St\"ockinger,
{\em Eur.\ Phys.\ J.}{\bf\ C} {\bf 20} (2001) 105
[arXiv:hep-ph/0103009];\\
I.~Fischer, W.~Hollik, M.~Roth and D.~St\"ockinger,
{\em Phys.\ Rev.}{\bf\ D} {\bf 69} (2004) 015004
[arXiv:hep-ph/0310191].


\bibitem{SSTI1}
P.~L.~White,
{\em Class.\ Quant.\ Grav.}\  {\bf 9} (1992) 1663.

\bibitem{SSTI2}
N.~Maggiore, O.~Piguet and S.~Wolf,
{\em Nucl.\ Phys.}{\bf\ B }{\bf 458} (1996) 403
[Erratum-ibid.\ B {\bf 469} (1996) 513]
[arXiv:hep-th/9507045],
{\em Nucl.\ Phys.}{\bf\ B }{\bf 476} (1996) 329
[arXiv:hep-th/9604002].

\bibitem{SSTIus}
W.~Hollik, E.~Kraus and D.~St\"ockinger,
{\em Eur.\ Phys.\ J.}{\bf\ C }{\bf 23} (2002) 735
[arXiv:hep-ph/0007134];\\
W.~Hollik, E.~Kraus, M.~Roth, C.~Rupp, K.~Sibold and D.~St\"ockinger,
{\em Nucl.\ Phys.}{\bf\ B} {\bf 639} (2002) 3
[arXiv:hep-ph/0204350].


\bibitem{Siegel80}
W.~Siegel,
{\em Phys.\ Lett.}{\bf\ B }{\bf 94} (1980) 37.

\bibitem{HHW}
S.~Heinemeyer, W.~Hollik and G.~Weiglein,
arXiv:hep-ph/0412214.

\bibitem{SPA}
P.~M.~Zerwas, Talk given at the ECFA {\em "Physics and Detectors for a
Linear Collider"} Workshop, Durham 1 - 4 September 2004; 
{\tt http://www.ippp.dur.ac.uk/ECFA04/program.html}.

\bibitem{BKNS}
W.~Beenakker, H.~Kuijf, W.~L.~van Neerven and J.~Smith,
{\em Phys.\ Rev.}{\bf\ D} {\bf 40} (1989) 54.

\bibitem{NS}
J.~Smith and W.~L.~van Neerven,
arXiv:hep-ph/0411357.

\bibitem{JJK}
I.~Jack, D.~R.~T.~Jones and A.~F.~Kord,
{\em Phys.\ Lett.}{\bf\ B }{\bf 579}, 180 (2004)
[arXiv:hep-ph/0308231].

\bibitem{JJReview}
I.~Jack and D.~R.~T.~Jones,
``Regularisation of supersymmetric theories'', 
in {\em Kane, G.L. (ed.): Perspectives on supersymmetry} 149-167;
[arXiv:hep-ph/9707278].

\bibitem{ACV}
L.~V.~Avdeev, G.~A.~Chochia and A.~A.~Vladimirov,
{\em Phys.\ Lett.}{\bf\ B} {\bf 105} (1981) 272;\\
L.~V.~Avdeev and A.~A.~Vladimirov,
{\em Nucl.\ Phys.}{\bf\ B }{\bf 219} (1983) 262.

\bibitem{Wilson}
K.~G.~Wilson,
{\em Phys.\ Rev.}{\bf\ D }{\bf 7} (1973) 2911.

\bibitem{Collins}
J.~Collins, {\em ``Renormalization''}, Cambridge Monographs on
Mathematical Physics.

\bibitem{BM}
               P.~Breitenlohner and D.~Maison,
               {\em Commun. Math. Phys.} {\bf 52} (1977) 11.


\bibitem{QAPBPHZ}
J.~H.~Lowenstein,
{\em Phys.\ Rev.}{\bf\ D} {\bf 4} (1971) 2281,
{\em Commun.\ Math.\ Phys.}\  {\bf 24} (1971) 1;
Y.~M.~Lam,
{\em Phys.\ Rev.}{\bf\ D} {\bf 6} (1972) 2145;
{\em Phys.\ Rev.}{\bf\ D} {\bf 7} (1973) 2943.

\bibitem{CJN80}
D.~M.~Capper, D.~R.~T.~Jones and P.~van Nieuwenhuizen,
{\em Nucl.\ Phys.}{\bf\ B} {\bf 167} (1980) 479.

\bibitem{BM85}
P.~Breitenlohner and D.~Maison,
``Renormalization Of Supersymmetric Yang-Mills Theories,''
in {\em Cambridge 1985, Proceedings, Supersymmetry and Its Applications}, 309-327.
\bibitem{betachecks}
S.~P.~Martin and M.~T.~Vaughn,
{\em Phys.\ Lett.}{\bf\ B} {\bf 318} (1993) 331
[arXiv:hep-ph/9308222];\\
I.~Jack, D.~R.~T.~Jones and C.~G.~North,
{\em Nucl.\ Phys.}{\bf\ B} {\bf 473} (1996) 308
[arXiv:hep-ph/9603386],
{\em Phys.\ Lett.}{\bf\ B} {\bf 386} (1996) 138
[arXiv:hep-ph/9606323],
{\em Nucl.\ Phys.}{\bf\ B} {\bf 486} (1997) 479
[arXiv:hep-ph/9609325].

\bibitem{SiboldSusyTrans}
C.~Rupp, R.~Scharf and K.~Sibold,
{\em Nucl.\ Phys.}\ {\bf B} {\bf 612} (2001) 313
[arXiv:hep-th/0101165],
[arXiv:hep-th/0101153].

\bibitem{g-2}
S.~Heinemeyer, D.~St\"ockinger and G.~Weiglein,
{\em Nucl.\ Phys.}{\bf\ B }{\bf 699} (2004) 103
[arXiv:hep-ph/0405255].
D.~St\"ockinger,
{\em Nucl.\ Phys.\ Proc.\ Suppl.\  }{\bf 135} (2004) 311
[arXiv:hep-ph/0406306].


\bibitem{Gamma5}
S.~A.~Larin,
{\em Phys.\ Lett.}{\bf \ B} {\bf 303} (1993) 113
[arXiv:hep-ph/9302240];
T.~L.~Trueman,
{\em Z.\ Phys.}{\bf\ C }{\bf 69} (1996) 525
[arXiv:hep-ph/9504315].


\end{thebibliography}
\end{document}